\documentclass[aps,prl,reprint,groupedaddress]{revtex4-2}
\usepackage{amsmath}
\usepackage{amsfonts}			
\usepackage{graphicx}
\usepackage{xcolor}

\usepackage{hyperref}  
\usepackage{amsmath,bm}  

\begin{document}


\title{Locally Resonant Metagrating by Elastic Impedance Modulation}

%
\author{Liyun Cao*, Sheng Wan, Badreddine Assouar*}

\affiliation{Université de Lorraine, CNRS, Institut Jean Lamour, Nancy 54000, France.}

\date{\today}

\begin{abstract}
The optical and acoustic metagratings have addressed the limitations of low-efficiency wave manipulation and high-complexity fabrication of metamaterials and metasurfaces. In this research, we introduce the concept of elastic metagrating and present the theoretical and experimental demonstration of locally resonant elastic metagrating (LREM). Remarkably, the LREM, with dimensions two orders of magnitude smaller than the relevant wavelength, overcomes the size limitations of conventional metagratings and offers a unique design paradigm for highly efficient wave manipulation with an extremely compact structure in elastic wave systems. Based on a distinctive elastic impedance engineering with hybridization of intrinsic evanescent waves, the proposed LREM achieves wide-angle perfect absorption. This tackles a fundamental challenge faced by all elastic metastructures designed for wave manipulation, which consists in the unavoidable vibration modes in finite structures hindering their implementations in real-world applications.
\end{abstract}

\maketitle

Elastic wave manipulation in functional devices, typically in plate-like chip devices, has shown significant potential for various applications, such as high signal-to-noise information processing \cite{RN11300,RN11484,RN11469}, sensing \cite{RN11299,RN11490}, and wave-matter interaction for future quantum networks \cite{RN11376,RN11379}. Elastic metamaterials and metasurfaces (MMs) \cite{RN11298,RN11033,RN11196,RN5980,RN11359,RN11525,RN10141}, engineered microstructures analogous to optical \cite{RN4309,RN9408} and acoustic \cite{RN11493,RN11491,RN9699,Donda_2021} MMs, have achieved remarkable advancements in elastic wave manipulation, greatly surpassing the capabilities of natural materials. These engineered elastic structures have even enabled the experimental realization of some wave physical phenomena that are inherently challenging to achieve in quantum and electromagnetic systems, such as extreme curvature wormhole \cite{RN11449}, below-diffraction-limit focused imaging \cite{RN11451}, and phase-shift-free cloaking \cite{RN11455,RN11473}. However, elastic MMs face inherent challenges due to the fact that both the elastic MMs and manipulated elastic waves must share the same finite carriers, for instance, in a typical plate-like chip carrier illustrated in Fig. 1(a). Indeed, the manipulated elastic waves undergo continuous reflection at the mechanical boundaries of the carrier, resulting in the formation of undesired standing wave fields, i.e., vibration modes. These vibration modes hamper any wave manipulation of elastic MMs within the carrier structure. The conventional attempt to address this issue in some wave-manipulation experiments is coating bulky loss appendages with low-efficiency absorption in all boundaries \cite{RN11300,RN11451,RN5980,RN11473,RN11475}. However, this only partially mitigates the adverse effects of the vibration modes, leading to a real degradation of manipulation performance. Moreover, implementing such bulky appendages in compact real-world devices is often impractical. So far, effective suppression of vibration modes, especially those originating from arbitrary mechanical boundaries, using compact appendages, remains a significant and unresolved challenge.

The emergence of optical and acoustic metagratings \cite{RN7964,RN11499,RN11500,RN11502} has recently garnered significant interest due to their ability to overcome the limitations associated with conventional MMs, such as low-efficiency wave manipulation and high-complexity fabrication. However, a weakness of these metagratings is their bulky nature, as their structural size is comparable to the wavelength, which is a result of nonlocal scattering characteristics \cite{RN11497,RN11498,RN11535}. On the other hand, locally resonant metamaterials \cite{RN11494,RN11501} use deep sub-wavelength resonant units to create resonance band gaps, leveraging the strong energy localization within these units. This property allows them to break the size limitations imposed on phononic crystals with Bragg band gaps. However, the application of these classical local resonators for modulating transmission of elastic wave energy has not been extensively explored, especially in terms of impedance modulation. Building upon this premise, our research presents the concept of metagratings for elastic waves and introduces the locally resonant elastic metagrating (LREM) by incorporating local resonant physics. This innovation overcomes the size limitations typically associated with conventional metagratings while retaining their significant advantages. Elaborate adjustments of the impedance of the resonant units within the hybridization of intrinsic evanescent fields preserve local energy conservation across various transverse momenta. This enables the LREM to efficiently manipulate elastic waves and achieve wide-angle perfect absorption for different mechanical boundaries. As a result, and as illustrated in Fig. 1(b), the ultra-compact LREM effectively suppresses vibration modes, addressing the unresolved challenges mentioned above.

\begin{figure}
\centering 
\includegraphics[height=6.5cm]{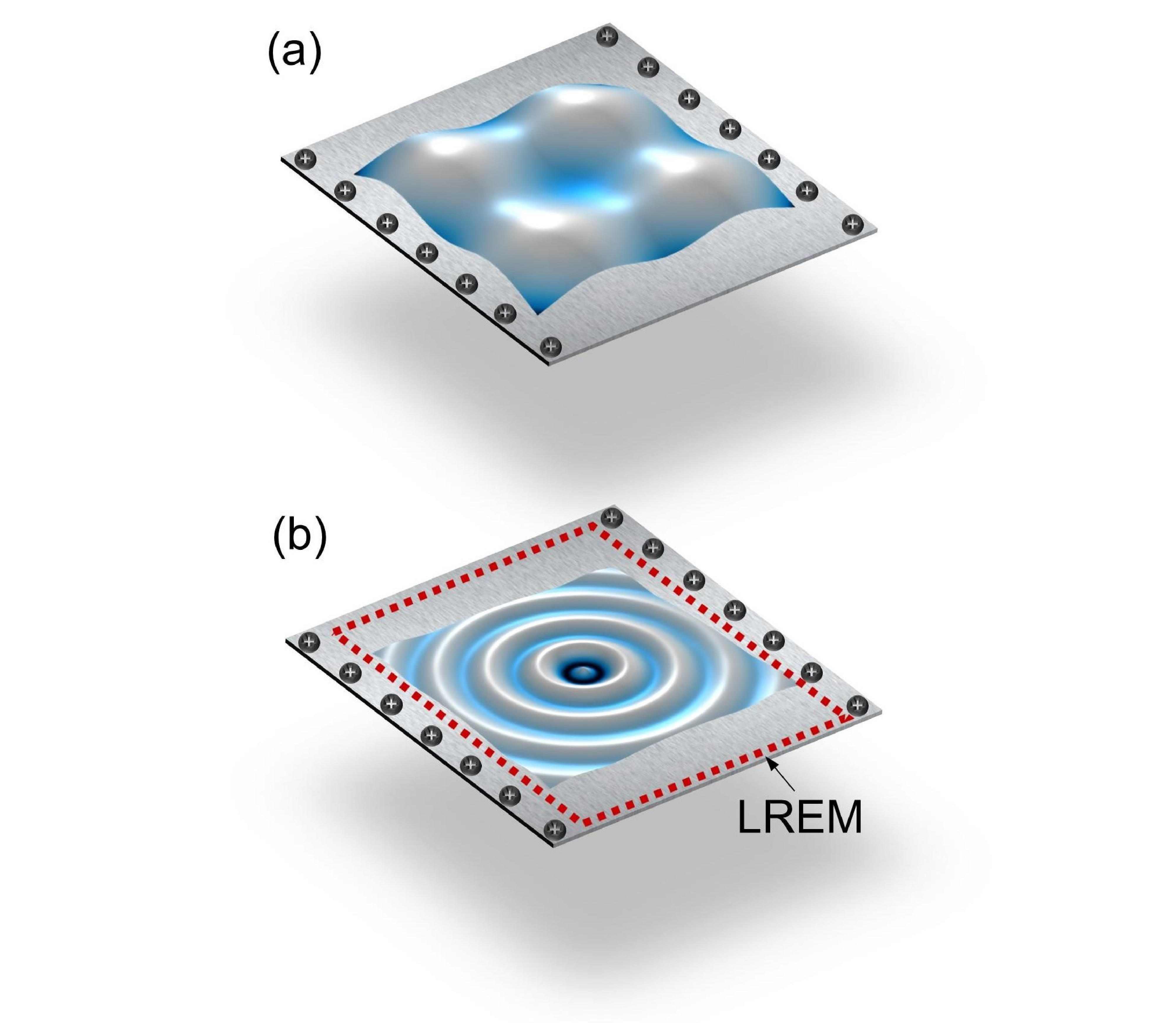} 
\caption{Schematics of the locally resonant elastic metagrating (LREM). (a) Illustration of a typical plate-like chip carrier with two fixed boundaries due to mounting screws and two free boundaries, showing a standing-wave vibration mode excited by a source at the center. (b) The engineered LREM effectively suppresses the vibration mode, enabling it to achieve a propagating wavefield originating from the point source. The LREM, characterized by its deep-subwavelength size and ultra-light mass, consists of an array of single units placed along all boundaries.} 
\label{1a}
\end{figure}
Fig. 2(a) presents the impedance model of the LREM unit, in which a universal lossy local resonator is attached to a plate-like background structure. The local resonator consists of the mass $\mathcal{M}$ and spring with the compliance $C_{\mathrm{m}}=\left [ \mathcal{K} \left ( 1+\eta \mathrm{i} \right )  \right ]^{-1}$. The loss factor $\eta$ describes its loss $R$, while the $\mathcal{K}$ denotes the spring stiffness. In acoustics, the impedance of the classical local resonator is defined as $Z_{\mathrm{a} }=  \mathrm{i} \left (\omega \mathcal{M}-\frac{1}{\omega C_{\mathrm{m} } }\right )$ \cite{RN11478,RN11480}. However, in mechanics with the continuous elastic media, we redefine the impedance as $Z_{\mathrm{m} }=  -\mathrm{i} \left (\omega C_{\mathrm{m}}-\frac{1}{\omega \mathcal{M}  }\right ) ^{-1}$, determined by the different equivalent mechanical circuit illustrated in Fig. 2(a). Within the classical Kirchhoff plate theory \cite{RN9991}, displacements $w$ of all propagating diffraction waves (flexural waves), scattered from the LREM to the background structure, obey the motion equation $\left ( \nabla ^{2}\nabla ^{2}+\varepsilon ^{4}  \frac{\partial^2}{\partial t^2}    \right ) \cdot w\left ( x,y,t \right ) =0$, where $\varepsilon =\left ( \rho h /D\right ) ^{1/4}$ is the propagation constant, and $D$ is the bending rigidity. $\rho$ and $h$ are the density and the thickness, respectively. The propagation properties of these elastic waves in the background structure are characterized by the force impedance of $Z_{V0}=\mathrm{i}D p k^3 /\omega$ (the ratio of shear force $V$ and velocity  $\partial w/ \partial t$) and the moment impedance of $Z_{M0}=\mathrm{i}D p k/\omega$ (the ratio of bending moment $M$ and angle velocity $\partial^{2}  w/ \partial x \partial t$), where $k=\varepsilon \sqrt{\omega }$ is flexural wavenumber and $\omega$ represents the angular frequency.
\begin{figure}
\centering 
\includegraphics[height=9.73cm]{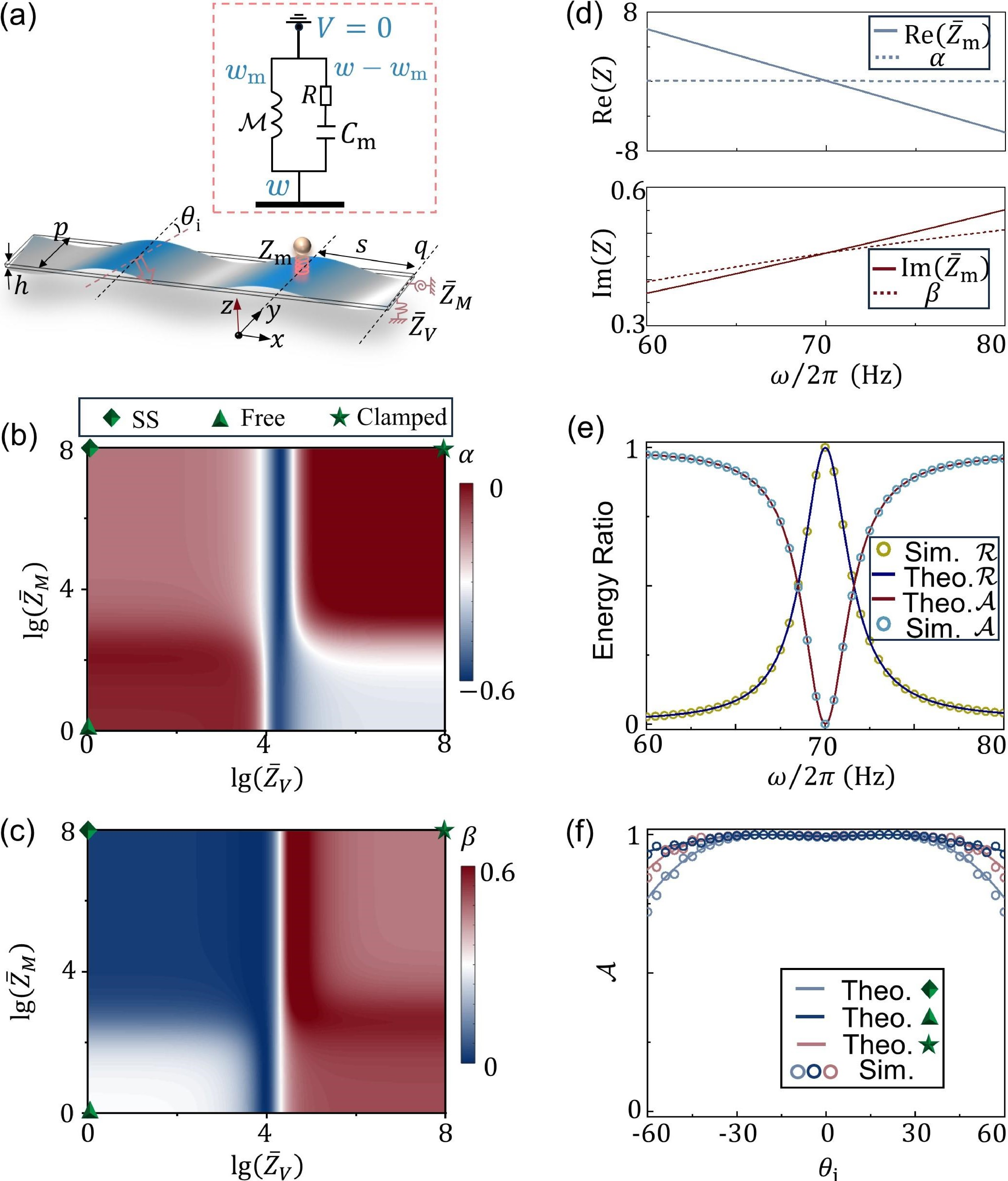} 
\caption{(a) Impedance model of the LREM unit. A mass-spring resonator, characterized by impedance $Z_{\mathrm{m} }$, is located at a distance $s$ from the arbitrary mechanical boundary $q$, which is characterized by boundary impedances $\bar{Z}_{V}$ and $\bar{Z}_{M}$. The left boundary is an infinity boundary. The equivalent mechanical circuit of the resonator is enclosed in the dashed box. The unit features a pair of period boundaries situated at the upper and lower long ends of the background structure in the $y$-axis direction, with a periodicity of $p$. As an example, the central incident angle $\theta _\mathrm{i}$ is set as 20 degrees. (b) and (c) Variations of the hybrid mechanical resistance $\alpha$ and reactance $\beta$ of the background structure as functions of  $\bar{Z}_{V}$ and $\bar{Z}_{M}$ at the central frequency of 70 Hz, respectively, with $s$ being 0.3 times the peak-absorption wavelength $\lambda_\mathrm{p}$. The green rhombus, triangles, and pentagrams represent the simply supported (SS) ($\bar{Z}_{V}\sim \infty$ and $\bar{Z}_{M} = 0$), free ($\bar{Z}_{V} = 0$ and $\bar{Z}_{M} = 0$), and clamped ($\bar{Z}_{V}\sim \infty$ and $\bar{Z}_{M} \sim \infty$) boundaries, respectively. (d) For a clamped boundary, the curves of $\mathrm{Re}\left(\bar{Z}_{\mathrm{m}}\right )$ and $\mathrm{Im}\left(\bar{Z}_{\mathrm{m}}\right )$ of the designed resonant unit intersect with the curves of $\alpha$ and $\beta$ at the central frequency, respectively. (e) Theoretical and simulated reflection $\mathcal{R}$ and absorption $\mathcal{A}$ spectra for the LREM. (f) Theoretical and simulated $\mathcal{A}$ of the LREM as a function of the incident angle $\theta _\mathrm{i}$, while the LREM is designed for the central incident angle of 20 degrees. The wide-angle absorption can be achieved for various boundaries, such as the SS boundary with $s=0.3\lambda_\mathrm{p}$, clamped boundary with $s=0.3\lambda_\mathrm{p}$, and free boundary with $s=0$, respectively.} 
\label{1a} 
\end{figure}

Mechanical boundaries are inherent aspects of practical devices, as indicated by the dotted line $q$ in Fig. 2(a). We harness these boundaries within the impedance model to manipulate elastic waves effectively. Notably, the model exhibits universality and can be applied to arbitrary mechanical boundaries, such as classical free, simply supported (SS), and clamped boundaries, as shown in Figs. 2(b) and 2(c).  Different from optical and acoustic boundaries described by single-degree-of-freedom impedance, we need to characterize mechanical boundaries by a two-degrees-of-freedom impedance vector $\mathcal{T } =\left [  \begin{matrix}
 \bar{Z}_V  & \bar{Z}_M & \bar{Z}_V \cdot \bar{Z}_M & 1
\end{matrix}\right ]^{T} $, where $\bar{Z}_V= Z_{V}^{b} /Z_{V0}$ and $\bar{Z}_M= Z_{M}^{b} /Z_{M0}$ are the dimensionless force and moment boundary impedances, respectively. We, for the first time, have established a robust experimental model to accurately measure  $\mathcal{T}$ for arbitrary mechanical boundaries (see Supplemental Material, S5 \cite{SM001}). This serves as a foundation to experimentally explore elastic wave physics within various mechanical boundaries.

The LREM scatters the incident wave with an incidence angle $\theta _\mathrm{i}$ into $N$ diffraction waves. These waves subsequently propagate into the far field, and some are transmitted to the mechanical boundary. Each of these propagating diffraction waves is governed by the above fourth-order motion equation, a distinction from the second-order equations encountered in optics and acoustics. This fourth-order equation yields four solutions for $x$-component momentum \cite{RN9991,RN11397}. These solutions encompass two real values $^{\pm  } k_{nx}=\pm \sqrt{k^2-k_{ny}^{2} } $ and two additional purely imaginary ones $^{\pm  } \tilde{k} _{nx}= \pm \mathrm{i} \sqrt{k^2+k_{ny}^{2}}$, with $n$ signifying the $n^{\mathrm{th} }$ propagating diffraction wave. These  $x$-component momentums are determined by transverse momentums $k_{ny}$. The intrinsic evanescent fields $^{\pm } {\varphi }_{n\mathrm{e}} \propto e^{-\mathrm{i} \cdot k_{ny}\cdot y }\cdot e^{\pm \mathrm{i} \cdot ^{\pm }  \tilde{k}_{nx}\cdot x }$ are associated with these imaginary solutions, and they hybridize with propagating diffraction fields $^{\pm } {\varphi }_{n\mathrm{p}} \propto e^{-\mathrm{i} \cdot k_{ny}\cdot y }\cdot e^{\pm \mathrm{i} \cdot ^{\pm } {k}_{nx}\cdot x }$ transmitted at the resonant units and reflected from the mechanical boundaries. To effectively absorb the incident wave energy, the periodicity $p$ of the LREM is deliberately chosen to be significantly smaller than half of the relevant wavelength. This ensures that only the $0^{\mathrm{th}}$ transverse momentum exists, based on the diffraction theorem \cite{RN10379}. Employing the mode-coupling method in elastic impedance system, we establish a theoretical model representing the intrinsic evanescent field $^{\pm } {\tilde{\varphi }}_{\mathrm{e}} \propto \left ( {\varphi}_{\mathrm{e}} \right )^{\pm 2} =\left ( e^{ \mathrm{i} \cdot  \tilde{k}_{0x}\cdot s } \right )^{\pm 2}$ hybridizing with propagating diffraction field $^{\pm } {\tilde{\varphi }}_{\mathrm{p}} \propto \left ( {\varphi}_{\mathrm{p}} \right )^{\pm 2} =\left ( e^{ \mathrm{i} \cdot  {k}_{0x}\cdot s } \right )^{\pm 2}$ (see Supplemental Material, S1 \cite{SM001}). The hybrid background impedance in the model can be succinctly expressed as:
\begin{equation}
Z_\mathrm{h}= \varsigma \cdot \left \langle \mathcal{T} \left | \bm{M}_ \mathrm{h}  \right | \mathcal{F} \right \rangle 
\tag{1}
\end{equation}
where $\varsigma =\mathrm{sinc} \left ( k_{0y}p/2 \right ) ^{-1}$ represents the transverse momentum constant. $\mathcal{F}  =\left [ \begin{matrix}{\varphi_\mathrm{p}}^2 
  & {\varphi_\mathrm{e}}^2  & {\varphi_\mathrm{p} \varphi_\mathrm{e}} & 1
\end{matrix} \right ]^{T} $ denotes the transfer vector between the resonant unit and the mechanical boundary, while $\bm{M} _\mathrm{h} =\frac{\xi }{\left \langle \mu_0  | \mathcal{T}   \right \rangle } $ represents the hybridization operator. Additionally, $\xi$ and $\mu_0$ denote the 4×4 hybridization coefficient matrix and boundary coefficient vector, respectively. These values within $\xi$ and $\mu_0$ are all determined by the single variable Poisson's ratio $\upsilon $ for a given incident angle $\theta _\mathrm{i}$, demonstrating the inherent hybridization characteristics of the elastic wave system (see details in Supplemental Material, S1 \cite{SM001}).

By leveraging the impedance modulation between $Z_\mathrm{h}$ and $Z_\mathrm{m}$, the reflected energy from the LREM can be effectively manipulated. This impedance-modulation manipulation can be characterized by the reflection coefficient:
\begin{equation}
r=e^{-2\mathrm{i} k_{0x}s} \frac{\bar{Z}_{\mathrm{m}}-Z_{\mathrm{h}}}{\bar{Z}_{\mathrm{m}}Z_{1}-Z_{2}} 
\tag{2}
\end{equation}
where $\bar{Z} _\mathrm{m}=Z_{V0}/Z_\mathrm{m}$  represents the dimensionless impedance of the resonant unit, while $Z_1=-\frac{\left \langle {{\mu}_0}^\ast   | \mathcal{T}   \right \rangle }{\left \langle \mu_0  | \mathcal{T}   \right \rangle}$ and $Z_2=\varsigma \cdot \frac{\left \langle \mathcal{T}  \left | \xi^\ast   \right | \mathcal{F}^{\ast}  \right \rangle  }{\left \langle \mu_0  | \mathcal{T}   \right \rangle} $ denote the system impedances, with the superscripts $\ast$ indicating complex conjugation. For a vertically incident wave ($\theta _\mathrm{i}=0$), if the boundary becomes a specific free boundary ($\bar{Z}_{V} = \bar{Z}_{M} = 0$), and both the evanescent field $\varphi_\mathrm{e}$ and the distance $s$ vanish, this elastic impedance system can be simplified to resemble an acoustic or optic one, with the conventional reflection coefficient of $r=\frac{Z_\mathrm{t}-1}{Z_\mathrm{t}+1}$. Particularly, when $\bar{Z} _{\mathrm{m}}$ equals $Z_\mathrm{h}$, indicating impedance matching, the reflection coefficient $r$ becomes 0, resulting in perfect energy absorption. The impedance matching occurs when the parameters of the resonant unit satisfy the simple equations:
\begin{equation} 
\mathcal{M} = \frac{\mathrm{Im}\left ( Z_{V0} \right ) \cdot \eta  }{\omega \left ( \alpha \eta +\beta  \right ) },  \ \ \ \ \ \  \mathcal{K} = \frac{\mathrm{Im}\left ( Z_{V0} \right ) \cdot \omega \eta  }{\beta  }   
\tag{3}
\end{equation}
where $\alpha$ and  $\beta$ represent the real and imaginary components of $Z_\mathrm{h}$, and they are respectively referred to the hybrid mechanical resistance and reactance of the background structure.

\begin{figure}
\centering 
\includegraphics[height=9.58cm]{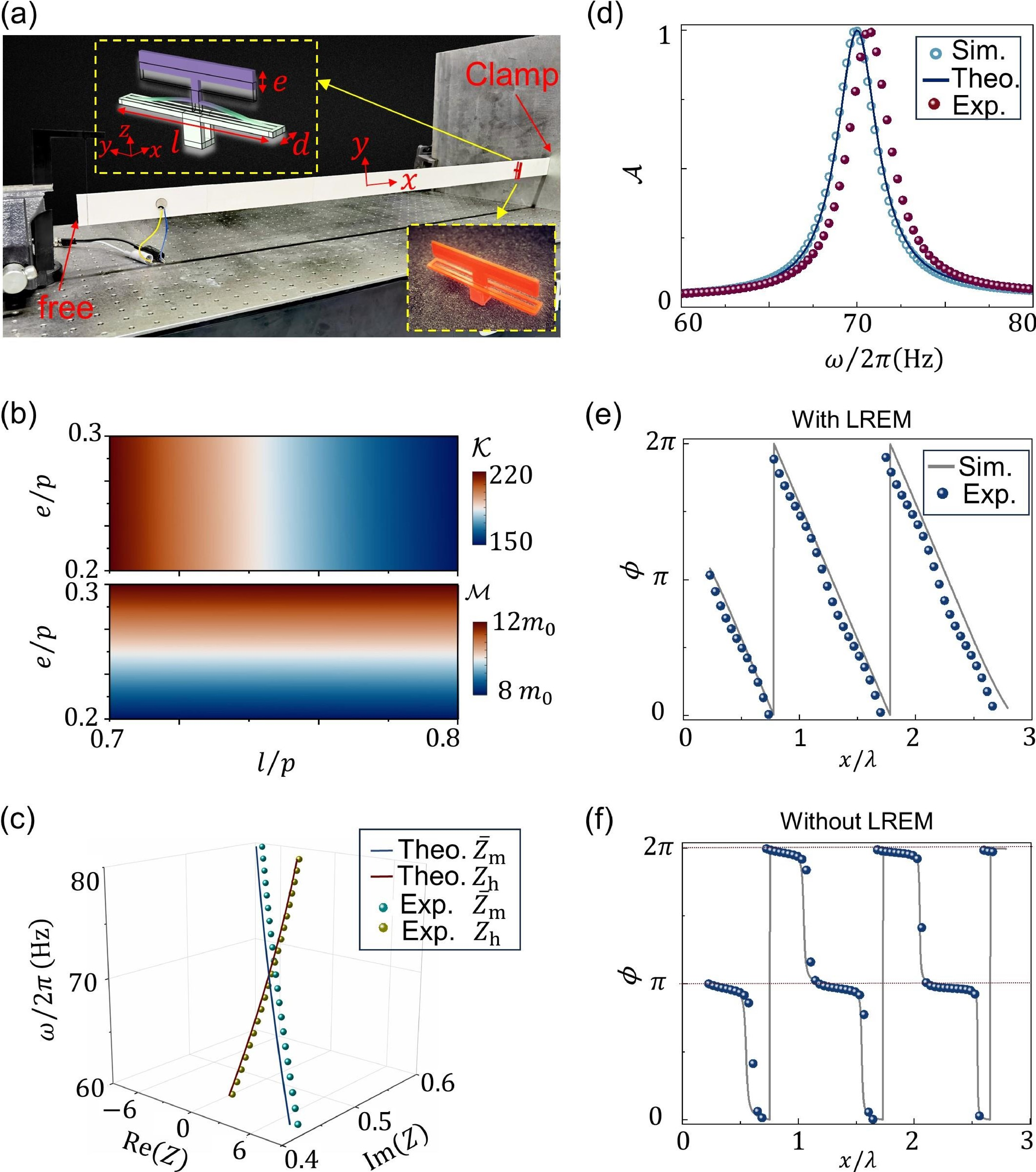} 
\caption{(a) Experiment setup for the resonant unit within a finite beam-like structure, serving as an equivalent representation of the LREM within a plate-like structure. The resonant unit is shown in the lower right corner, with its simulated eigenmode shown in the upper left corner. (b) Independent variations of stiffness $\mathcal{K}$ and mass $\mathcal{M}$ of the unit as functions of its geometrical sizes $l$ and $e$, respectively, with $m_0=1×10^{-4}$ Kg. (c) Experimental and theoretical curves of $Z_\mathrm{h}$ intersecting the corresponding curves of $\bar{Z}_\mathrm{m}$ at the central frequency of 70 Hz. (d) Experimental, theoretical, and simulated absorption spectra of the LREM. (e) and (f) Experimental and simulated phase fields within the background structure, both with and without LREM, along the $x$-axis.} 
\label{1a} 
\end{figure}

In this representative case, we consider an incident wave propagating within an aluminum-alloy background structure at the central frequency of 70 Hz, characterized by a central incident angle of 20 degrees. According to Eq. (1), different combinations of $\bar{Z}_V$ and $\bar{Z}_M$ result in distinct values for $\alpha$ and $\beta$, as illustrated in Figs. 2(b) and 2(c). This clearly demonstrates that mechanical boundary configuration is a critical degree of freedom for wavefield manipulation. In this case, the right boundary of the background structure is configured as a clamped boundary, one of the most common types. Referring to Figs. 2(b) and 2(c), we extract $\alpha$ and $\beta$ values. Subsequently, we calculate the corresponding $\mathcal{M}$ and $\mathcal{K}$ values of the resonant unit using Eq. (3). As shown in Fig. 2(d), $\mathrm{Re} \left ( \bar{Z}_\mathrm{m}  \right )$ and $\mathrm{Im} \left ( \bar{Z}_\mathrm{m}  \right ) $ of the designed unit match $\alpha$ and $\beta$ values well at the central frequency, respectively. In Fig. 2(e), the theoretical and simulated reflection $\mathcal{R}=\left | r \right | ^2$ and absorption $\mathcal{A}=1-\mathcal{R}$ spectra of the designed LREM exhibit perfect zero reflection. It is important to highlight that the LREM achieves an extremely light mass, compared with the background mass corresponding to the peak-absorption wavelength $\lambda _{\mathrm{p} } $. The specific mass ratio $\delta =ph\rho \lambda _{\mathrm{p} } /\mathcal{M} $ reaches an impressive value of 139. Remarkably, as depicted in Fig. 2(f), the LREM achieves perfect absorption over an ultra-wide range of incident angles, despite being designed at the central incident angle of 20 degrees. The notable efficiency is attributed to a wider range of incident angles where impedance matches are effectively compensated by the hybridization within the intrinsic evanescent field, which is in contrast to the scalar acoustic scenarios (see Supplemental Material, S3 \cite{SM001}).

For experimental convenience, we demonstrate the manipulation of a vertically incident wave using the resonant unit within a simplified beam-like background structure, which can serve as an equivalent representation of LREM within a two-dimensional plate-like structure (Fig. S5). To maintain generality, we set the left and right boundaries of this finite background structure as free and clamped boundaries, respectively, as shown in Fig. 3(a). The resonant unit is designed following metamaterial engineering. Its thin sub-beam, acting as an equivalent spring, is indicated by the largest deformation part within the simulated eigenmodes in Fig. 3(a). Independently varying stiffness $\mathcal{K}$ and mass $\mathcal{M}$ of the unit can be achieved by adjusting its geometric dimensions, represented by $l$ and $e$, respectively, as shown in Fig. 3(b). Note that our theoretical framework is not limited to this specific unit structure. It can be applied to various resonant models, including rubber-coated lead sphere \cite{RN11494}, magnet oscillators \cite{RN11523}, and spiral springs\cite{RN11522}.

Within this experimental setup presented in Fig. 3(a), a piezoelectric transducer (PZT) is positioned on the left side of the background structure, serving to generate an incident wave resembling a plane wave. We have measured the boundary impedance vector  $\mathcal{T}$ using a PSV-500 scanning laser Doppler vibrometer (SLDV) (see Supplemental Material, S5 \cite{SM001}). Subsequently, we obtain the experimental $Z_\mathrm{h}$, represented by the yellow spheres in Fig. 3(c). The experimental curves of $\bar{Z}_\mathrm{m}$ and $Z_\mathrm{h}$ nearly intersect at the central frequency, which agrees with our theoretical results. As shown in Fig. 3(d), the experimental, theoretical, and simulated absorption coefficients are all almost one in the vicinity of the central frequency. It is worth noting that the peak-absorption wavelength $\lambda _\mathrm{p}$ is 108 times the thickness size $d$ of the LREM, and the corresponding mass ratio $\delta $ is as high as 131. To further understand the LREM, we have conducted measurements of the phase field within this finite background structure along the $x$-direction using SLDV. As shown in Fig. 3(e), the phase field exhibits a periodic linear distribution spanning from 0 to $2\pi$. In contrast, the reference background structure without the LREM (Fig. 3(f)) shows a phase distribution that alternates approximately between the two values of $\pi$ and $2\pi$, indicating a typical standing-wave vibration mode. This contrast serves as clear evidence that the LREM effectively suppresses vibration modes within the finite background structure, owing to its exceptional absorption capabilities and robust out-of-plane polarization resonance. It is important to highlight that the LREMs fundamentally differ from conventional dynamic vibration absorbers (DVA) \cite{RN11524}, which fall short in suppressing these vibration modes (see Supplemental Material, S4 \cite{SM001}).

\begin{figure} 
\centering 
\includegraphics[height=9.1cm]{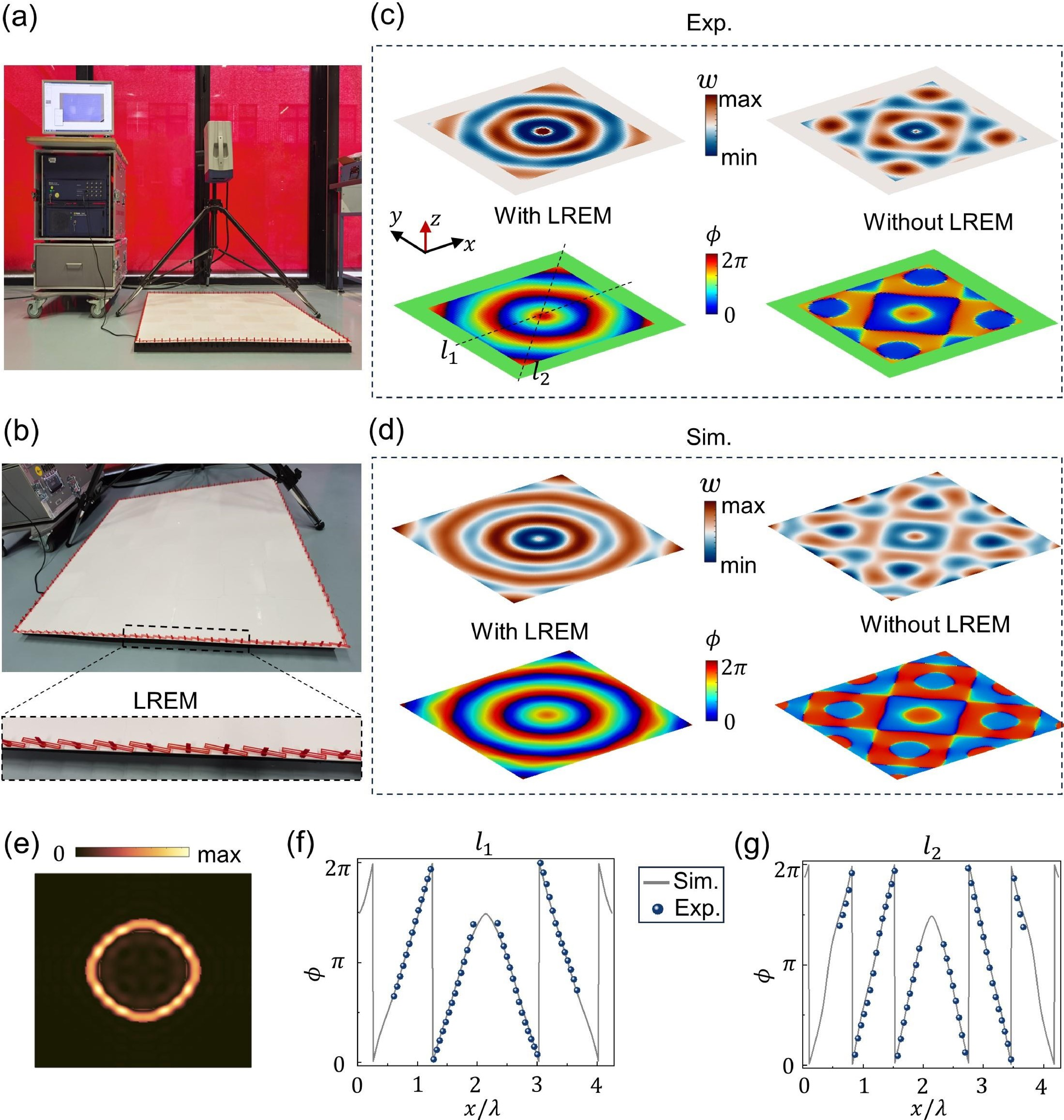} 
\caption{(a) Experimental setup for LREM within a finite plate-like structure. (b) The compact LREM, consisting of a periodic array of red resonant units along the boundaries of the white plate-like background structure. (c) Experimental displacement $w$ and phase fields $\phi$ within the background structure, both with and without the LREM. (d) Corresponding simulated displacement and phase fields. (e) 2D fast Fourier transform (FFT) of the experimental displacement field within the background structure with the LREM. (f) and (g) Experimental and simulated phase fields along lines $l_{1}$ and $l_{2}$, indicated in (c).} 
\label{1a} 
\end{figure}

Furthermore, as illustrated in Fig. 4(a), we have conducted experiments to demonstrate the capability of the LREM in effectively suppressing vibration modes within a finite plate-like background structure, leveraging its wide-angle absorption characteristics. To verify the universality of the LREM, we have altered four mechanical boundaries of the background structure to free boundaries at the central frequency of 100 Hz. According to Eqs. (1) and (3), we have modified the geometry of the resonant units. In Fig. 4(c), the experimental displacement field within the background structure equipped with the LREM, as measured by SLDV, exhibits a point-source field excited by a PZT at the center. Additionally, the phase field exhibits a periodic linear distribution spanning from 0 to $2\pi$ along the radial propagation direction, as evident in Figs. 4(f) and 4(g). In contrast, the reference background structure demonstrates a typical standing-wave vibration mode. All simulated results in Fig. 4(d) are consistent with the experimental results. In Fig. 4(e), we present the 2D fast Fourier transform (FFT) of the experimental displacement field in Fig. 4(c). The highlighted ring-shaped area confirms that the presence of a point-source wavefield within the engineered finite structure remains unaffected by the vibration mode.

In conclusion, this research has theoretically and experimentally demonstrated the general concept of LREM with a deep sub-wavelength size, ultra-light mass, and exceptionally simple configuration, which is a remarkable achievement not previously realized in the realm of elastic metamaterials and metasurfaces (see Supplemental Material, S9 \cite{SM001}). The proposed LREM effectively addresses a persistent challenge faced by conventional metastructures introduced by vibration modes in real devices.
Interestingly, the innovative concept of the LREM can be extended to manipulate reflection wavefronts with unitary efficiency, employing a similar impedance modulation. Notably, the LREM can achieve unitary retroreflection even at an extremely incident angle of 75 degrees, with an impressive mass ratio $\delta $ reaching 1161 (see Supplemental Material, S2 \cite{SM001}). This is unattainable through conventional elastic metamaterials and metasurfaces \cite{RN5980,RN11536,RN11538,RN11537}. Beyond this, by locally adjusting the period  \cite{RN11496,RN7964}, the LREM enables elastic wave steering in various directions, opening possibilities for applications such as focusing, holography, and other wave transformations. More than that, this research also paves the way for further investigations into impedance-engineered wave manipulation within elastic wave systems, particularly for various boundary impedances, which have recently shown promise in the acoustics \cite{RN11487}.

\vspace{0.5cm}

\begin{acknowledgments}
This work was supported by the Air Force Office of Scientific Research under award number FA9550-18-1-7021, and  la Région Grand Est.
\end{acknowledgments}

\vspace{0.1cm}
*Corresponding author: liyun.cao@univ-lorraine.fr, badreddine.assouar@univ-lorraine.fr

\end{document}